\documentclass[12pt]{article}
\usepackage{graphicx}
\renewcommand{\baselinestretch}{1.}
\begin{document}
\begin{center}
{\Large\bf Transverse momentum and pseudorapidity distributions of
final-state particles and spatial structure pictures of
interacting system in $p$-Pb collisions at \(\sqrt{s_{NN}}=5.02\)
TeV}

\vskip1.0cm

Fu-Hu Liu$^{a,}${\footnote{E-mail: fuhuliu@163.com;
fuhuliu@sxu.edu.cn}}, Hua-Rong Wei$^{a}$, and Roy A.
Lacey$^{b,c,}${\footnote{E-mail: Roy.Lacey@Stonybrook.edu}}

{\small\it $^a$Institute of Theoretical Physics, Shanxi
University, Taiyuan, Shanxi 030006, China

$^b$Department of Chemistry, Stony Brook University, Stony Brook,
NY 11794-3400, USA

$^c$Department of Physics and Astronomy, Stony Brook University,
Stony Brook, NY 11794-3800, USA}
\end{center}

\vskip1.0cm

{\bf Abstract:} The transverse momentum and pseudorapidity
distributions of final-state particles produced in proton-lead
($p$-Pb) collisions at center-of-mass energy per nucleon pair
$\sqrt{s_{NN}}=5.02$ TeV are studied in the framework of a
multisource thermal model. Experimental results measured by the
ALICE and CMS Collaborations are described by the Tsallis
transverse momentum distribution and the two-cylinder
pseudorapidity distribution. Based on the parameter values
extracted from the transverse momentum and pseudorapidity
distributions, some other quantities are extracted. Then, the
structure pictures of interacting system at the stage of kinetic
freeze-out in some spaces are obtained.
\\

\\

\\

{\section{Introduction}}

High energy collisions are an important research field in modern
physics. Since 2000, the Relativistic Heavy Ion Collider (RHIC)
has been opening a new era for the collisions, which boosts
superlatively the center-of-mass energy per nucleon pair
($\sqrt{s_{NN}}$) to 200 GeV [1--8]. In 2008, the Large Hadron
Collider (LHC) ran firstly. Presently, the LHC carries out
proton-proton ($pp$), proton-lead ($p$-Pb), and lead-lead (Pb-Pb)
collisions at different TeV [9--16]. Such high energy collisions
give us a chance to study not only the properties of quark-gluon
plasma (QGP) and other new physics but also the particle spectra
of different distributions and related effects.

A high density and high temperature location is expected to form
in high energy nucleus-nucleus collisions which provides a method
to create QGP matter and to produce multiple particles. As an
input quantity in nucleus-nucleus collisions, $pp$ collision is
very useful to understand the interacting mechanisms of
nucleus-nucleus collisions. For the purpose of understanding the
nuclear effect and whole collision process, one also needs to
study the intervenient proton-nucleus collisions which is the
topic of the present work.

Transverse momentum distributions are very important to understand
the degrees of transverse excitation and non-equilibrium of
interacting system. Different distribution forms are used to
describe experimental transverse momentum distributions.
Generally, for example, we need a two- or multi-Boltzmann
(Fermi-Dirac) distribution or other distributions to describe wide
transverse momentum distributions. This renders a two- or
multi-temperature emission picture which is in fact to fall into
the framework of a multisource thermal model [17--19]. This also
means that the interacting system has temperature changes
(fluctuations) from a temperature to another one, which can be
described by the Tsallis statistics [20--29].

Pseudorapidity (or rapidity) distributions are very important to
understand the longitudinal extension of interacting system and
nuclear stopping in heavy ion collisions. A pseudorapidity
distribution in whole phase space contains not only the
contribution of violent interacting components but also the
contribution of leading nucleons. A pseudorapidity distribution in
central region contains only the contribution of violent
interacting components. A combined analysis on transverse momentum
and pseudorapidity distributions can provide abundant information
on processes of high energy collisions.

In this paper, in the framework of the multisource thermal model
[17--19], we use the Tsallis statistics [20--29] to describe the
transverse momentum distributions of final-state particles
produced in $p$-Pb collisions with different centrality intervals
at $\sqrt{s_{NN}}=5.02$ TeV. Meanwhile, the pseudorapidity
distribution of charged particles produced in
non-single-diffractive (NSD) $p$-Pb collisions at the same energy
is studied by using the same model. The calculated results are
compared with the experimental data of the ALICE [30, 31] and CMS
Collaborations [32]. Some quantities are then extracted from the
comparisons, and structure pictures of interacting system in some
spaces are obtained due to the extractions.
\\

{\section{The model and calculation}}

In the multisource thermal model [17--19], we assume that many
emission sources are formed in high energy collisions. Because
different interacting mechanisms exist in the collisions and
different event samples are measured in experiments, these sources
can be classified into a few groups. The sources in the same group
are assumed to stay at a local equilibrium state with a given
temperature. We can use different models to describe the local
equilibrium state. The final-state distribution is the result of a
two- or multi-temperature emission process. For example, we can
use a two- or multi-Boltzmann (Fermi-Dirac) distribution to
describe the transverse momentum distribution and to obtain two or
multiple temperatures. This means that the temperature is in fact
to have changes (fluctuations) from a local equilibrium state to
another one.

It is not a good choice for us to use the two or multiple
distributions to describe the transverse momentum spectra and
temperature fluctuations. We hope to use only one distribution to
describe uniformly the spectra, and only one temperature to
describe the mean effect of the temperature fluctuations. The good
candidate is the Tsallis statistics which is widely used in high
energy collisions [20--29]. The Tsallis statistics can be used not
only for the whole interacting system but also for the singular
source. Although one expects another set of parameters for the
latter one, we usually use the same set of parameters for both the
whole interacting system and the singular source.

According to the Tsallis statistics, the unit-density function of
transverse momentum ($p_T$) and rapidity ($y$) for a given type of
particles is [20--25]
\begin{equation}
\frac{d^2N}{dydp_T}=C_Tp_T\sqrt{p_T^2+m_0^2}\cosh y
\bigg[1+\frac{q-1}{T} \sqrt{p_T^2+m_0^2} \cosh y
\bigg]^{-\frac{q}{q-1}},
\end{equation}
where $N$ is the number of particles, $C_T=gV/(2\pi)^2$ is the
normalization constant, $g$ is the degeneracy factor, $V$ is the
volume, $T$ is the (average) effective temperature over
fluctuations in different groups, $q$ is the factor (entropy
index) to characterize the degree of non-equilibrium among
different groups, and $m_0$ is the rest mass of the considered
particle.

To give solely the transverse momentum distribution, we do an
integral for $y$ in Eq. (1). Then, we have the Tsallis transverse
momentum distribution to be
\begin{equation}
f_{p_T}(p_T)=\frac{1}{N}\frac{dN}{dp_T}=C_0 p_T\sqrt{p_T^2+m_0^2}
\int_{y_{\min}}^{y_{\max}} \cosh y
\bigg[1+\frac{q-1}{T}\sqrt{p_T^2+m_0^2} \cosh y
\bigg]^{-\frac{1}{q-1}} dy
\end{equation}
which uses the approximation expression $1/(q-1)$ [21--26] instead
of $q/(q-1)$ in the power index because $q$ is very close to 1,
where $C_0$ denotes the normalization constant which is
proportional to the volume, $y_{\max}$ denotes the maximum
rapidity, and $y_{\min}$ denotes the minimum rapidity.

On the pseudorapidity (or rapidity) distribution, in the
laboratory or center-of-mass reference frame, these multiple
sources are assumed to distribute at different rapidities $y_x$ in
the rapidity space. Because of different origins, these sources
are expected to form a target cylinder in (left) rapidity interval
$[y_{T\rm min},y_{T\rm max}]$ and a projectile cylinder in (right)
rapidity interval $[y_{P\rm min},y_{P\rm max}]$. Meanwhile, a
leading target nucleon source and a leading projectile nucleon
source are expected to form at $y_x=y_T$ and $y_x=y_P$
respectively. In symmetric collisions such as $pp$ or Pb-Pb
collisions, we have $y_{T\rm min}=-y_{P\rm max}$, $y_{T\rm
max}=-y_{P\rm min}$, and $y_T=-y_P$. In asymmetric collisions such
as $p$-Pb collisions, we do not have these equations.

To describe a pseudorapidity distribution in whole phase space, we
need to consider the contributions of the two cylinders and the
two leading nucleon sources. For a pseudorapidity distribution in
central region, we need only the contribution of the two-cylinder.
That is to say, we can use the two-cylinder pseudorapidity
distribution to describe the spectrum in the central region. What
we do in the model is only to consider the each contribution of
target and projectile nuclei to the central fireball. It does not
mean that the central fireball can be completely divided into two
separate fireballs. In our analysis, we can obtain $T$ and $q$ for
a given type of particles by comparing Eq. (2) with experimental
transverse momentum distribution. The obtained values of $T$ and
$q$ can be used in the analysis of pseudorapidity distribution.

Based on Eq. (2), we can use the Monte Carlo method to obtain a
series of $p_T$. For the assumption of isotropic emission in the
source rest frame, the distributions of space angle $\theta'$ and
azimuthal angel $\phi'$ can be given by
$f_{\theta'}(\theta')=(1/2) \sin \theta'$ and
$f_{\phi'}(\phi')=1/(2\pi)$ respectively. Correspondingly, we can
obtain a series of $\theta'$ and $\phi'$ by the Monte Carlo
method. Then, we have the $x$-component of momentum
$p'_x=p_T\cos\phi'$, the $y$-component of momentum
$p'_y=p_T\sin\phi'$, the longitudinal momentum $p'_z=p_T/\tan
\theta'$, the momentum $p'=p_T/\sin \theta'$ or
$p'=\sqrt{p_T^2+p'^2_z}$, the energy $E'=\sqrt{p'^2+m_0^2}$, the
rapidity $y'\equiv (1/2)\ln [(E'+p'_z)/(E'-p'_z)]$, and so forth.

In the laboratory or center-of-mass reference frame, we have the
rapidity $y=y_x+y'$, the energy $E=\sqrt{p_T^2+m_0^2}\cosh y$, the
momentum $p=\sqrt{E^2-m_0^2}$, the longitudinal momentum
$p_z=\sqrt{p_T^2+m_0^2}\sinh y$, the $x$-component of momentum
$p_x=p'_x$, the $y$-component of momentum $p_y=p'_y$, the emission
angle $\theta=\arctan{(p_T/p_z)}$, the pseudorapidity
$\eta\equiv-\ln\tan(\theta/2)$, the transverse rapidity $y_T\equiv
(1/2)\ln [(E+p_T)/(E-p_T)]$, the rapidity in $ox$ axis direction
$y_1\equiv (1/2)\ln [(E+p_x)/(E-p_x)]$, the rapidity in $oy$ axis
direction $y_2\equiv (1/2)\ln [(E+p_y)/(E-p_y)]$, the velocity
$\beta=p/E=\sqrt{p_z^2+p_T^2}/E$, the longitudinal velocity
$\beta_z=p_z/E$, the transverse velocity $\beta_T=p_T/E$, the
$x$-component of velocity $\beta_x=p_x/E$, the $y$-component of
velocity $\beta_y=p_y/E$, and so forth. Particularly, let $t_0$
denote the time internal from initial collision to the stage of
kinetic freeze-out, we have the space coordinates of the
considered particle at the stage of kinetic freeze-out to be
$x=t_0\beta_T\cos\phi'$, $y=t_0\beta_T\sin\phi'$,
$r_T=t_0\beta_T$, and $z=t_0\beta_z$.

We would like to point out that Eqs. (1) and (2) do not contain
the contribution of flow effect. For a local equilibrium state,
the final state distribution is contributed by the sum of thermal
motion and flow effect. In the case of neglecting the flow effect,
we may obtain a relative larger (effective) temperature. In the
case of considering the flow effect, we have experientially the
relations between quantities ($p'_x$, $p'_y$, $p'_z$, and $E'$) of
the thermal motion and quantities ($p_x$, $p_y$, and $p_z$) of the
thermal motion plus flow effect to be $p_x=(p'_x+\beta_x^{\rm
flow} E')/\sqrt{1-(\beta_x^{\rm flow})^2}$,
$p_y=(p'_y+\beta_y^{\rm flow} E')/\sqrt{1-(\beta_y^{\rm
flow})^2}$, and $p_z=(p'_z+\beta_z^{\rm flow}
E')/\sqrt{1-(\beta_z^{\rm flow})^2}$, while $\beta_x^{\rm flow}$,
$\beta_y^{\rm flow}$, and $\beta_z^{\rm flow}$ denote the $x$-,
$y$-, and $z$-components of flow velocity $\beta^{\rm flow}$,
respectively.

Generally, we can describe the thermal motion by the Tsallis
distribution and compare it with experimental transverse momentum
distribution to determine $\beta_x^{\rm flow}$ and $\beta_y^{\rm
flow}$. Then, $p_x$ and $p_y$ can be obtained in the case of
considering the flow effect. The azimuth $\phi=\arctan (p_y/p_x)$,
the directed flow $v_1=\cos \phi$, the elliptic flow
$v_2=\cos(2\phi)$, and the higher flow $v_n=\cos(n\phi)$ can be
naturally obtained. In the calculation, the directed transverse
motion of the emission source is considered in $\beta_x^{\rm
flow}$ and $\beta_y^{\rm flow}$. The directed longitudinal motion
of the emission source can be considered in $\beta_z^{\rm flow}$
or $y$.
\\

{\section{Comparisons, discussions, and extractions}}

Fig. 1 presents the transverse momentum distributions of (a)
$\pi^+$, $K^+$, and $p$, as well as (b) $\pi^-$, $K^-$, and $\bar
p$ produced in $p$-Pb collisions at $\sqrt{s_{NN}}=5.02$ TeV,
where $N_{EV}$ on the axis denotes the number of events. The
symbols represent the experimental data of the CMS Collaboration
measured in the rapidiy range $|y|<1$ [32]. The curves are our
results calculated by using the Tsallis transverse momentum
distribution [Eq. (2)]. The values of free parameters ($T$ and
$q$), normalization constant ($C_0$), and $\chi^2$ per degree of
freedom ($\chi^2$/dof) are given in Table 1. One can see that the
Tsallis distribution describes the experimental data of the
considered particles in $p$-Pb collisions at $\sqrt{s_{NN}}=5.02$
TeV. The temperature parameter $T$ increases and the
non-equilibrium degree parameter $q$ decreases with increase of
the particle mass, which maybe reflect non-simultaneous
productions of different types of particles. The normalization
constant decreases with increase of the particle mass, which will
be seen to conflict with the results discussed in Fig. 2.

Fig. 2 shows the transverse momentum distributions of (a)
$\pi^{\pm}$, (b) $K^{\pm}$, (c) $p+\bar p$, and (d) $\Lambda+\bar
\Lambda$ produced in $p$-Pb collisions at $\sqrt{s_{NN}}=5.02$
TeV. The symbols represent the experimental data of the ALICE
Collaboration measured in $0<y<0.5$ and different centrality ($C$)
intervals [30]. The curves are our results calculated by using the
Tsallis transverse momentum distribution [Eq. (2)]. For the
purpose of clearness, the results for different $C$ intervals are
scaled by different amounts shown in the panels. The values of
free parameters, normalization constant, and $\chi^2$/dof are
given in Table 1. Once more, the Tsallis distribution describes
the experimental data of the considered particles in $p$-Pb
collisions with different centrality intervals. The parameter $T$
decreases and the parameter $q$ increases with increase of the
centrality percentage, while $T$ and $q$ increases and decreases
with increase of the particle mass respectively. The dependence
trend of the normalization constant on the particle mass is
different from that in Fig. 1, although the normalization constant
increases with increase of the centrality percentage.

In the above descriptions, the experimental data are available in
a narrow transverse momentum region [30, 32]. This situation
affects the extraction of parameters which may vary in a wide
transverse momentum region [33--35]. According to the ``soft +
hard" model [34, 35], the transverse momentum spectrum is
contributed by the sum of soft and hard parts (yields). The soft
yields come from the QGP (or usual hadronic matter) and the hard
yields come from jets. Generally, the soft yields contribute in a
narrow region, and the hard yields contribute in a wide region. In
the region considered in the present work, the contribution of
soft yields is main, and the contribution of hard yields can be
neglected. The parameter values obtained in the present work can
be regarded as the result of the soft yields. At the same time, to
fix flow velocity, we need azimuthal distribution and more other
data. For the purpose of convenience and as the first
approximation, we neglect flow velocity in the calculations in
Figs. 1 and 2. This treatment is consistent with the general
Tsallis statistics [29--29] and our previous work on Pb-Pb
collisions at 2.76 TeV [36].

From Table 1 we see that the relationship between $C_0$ and $C$ is
non-linear. Particularly, $C_0$ has a quick increase in peripheral
$p$-Pb collisions. Qualitatively, nuclear stopping is very small
in peripheral $p$-Pb collisions, then a very large longitudinal
extension of the interacting system can be obtained at very large
$C$. This results in a very long interacting region and then a
very large $C_0$ which is proportional to the volume. The
situation in central $p$-Pb collisions (with very small $C$) is
opposite. In peripheral Pb-Pb collisions, the increase of $C_0$ is
small because the interacting region (and then the volume) in
central Pb-Pb collisions is also large.

To see clearly the dependences of parameters $T$ and $q$ on $C$
and $m_0$ in NSD $p$-Pb collisions at $\sqrt{s_{NN}}=5.02$ TeV, we
present the relations (a) $T-C$ for different particles, (b) $q-C$
for different particles, (c) $T-m_0$ for different centrality
intervals, (d) $q-m_0$ for different centrality intervals, (e)
$T-q$ for different particles, and (f) $T-q$ for different
centrality intervals in Fig. 3. The symbols represent the
parameter values in Table 1 and the lines are our fitted results
by linear function
\begin{equation}
Y=aX+b,
\end{equation}
where $Y$ denotes $T$ or $q$, and $X$ denotes $C$, $m_0$, or $q$.
The units of $T$ and $m_0$ are GeV and GeV/$c^2$ respectively. The
values of coefficients ($a$ and $b$) and $\chi^2$/dof are given in
Table 2. One can see that $T$ decreases and $q$ increases with
increase of the centrality percentage. At the same time, $T$
increases and $q$ decreases with increase of the particle mass.
There is a negative correlation between $T$ and $q$ not only for
different centralities but also for different particles. All the
linear relationships presented in Fig. 3 are experiential results.

The dependence trends of $T$ on $m_0$, $q$ on $m_0$, and $T$ on
$q$ in $p$-Pb collisions in the present work are consistent with
those in Pb-Pb collisions in our previous work [36]. Although the
dependence trends of $T$ on $C$ and $q$ on $C$ in $p$-Pb
collisions are different from those in Pb-Pb collisions, they are
not incompatible. In Pb-Pb collisions, $T$ and $q$ do not show a
change with increase of $C$ from 0--5\% to 30--40\%, and $T$
decreases and $q$ increases with increase of $C$ from 40--50\% to
80--90\% [36]. The former case can be explained by the large
enough interacting region comparing with $p$-Pb collisions and
peripheral Pb-Pb collisions. The later one and $p$-Pb collisions
have no large enough interacting region. Large interacting region
results in high $T$ and low $q$ due to more energy deposits and
more scattering processes respectively.

Except for the possible non-simultaneous production, we have
another explanation on the dependence of $T$ on $m_0$. All the
temperatures obtained in the present work and our previous work
[36] are effective temperatures. If we consider the flow effect,
it is expected to obtain a uniform ``true" temperature of the
source for different particle emissions. This ``true" temperature
should be less than the weighted average of effective temperatures
for different particles due to subtracting the flow effect.
Because pions are absolutely the most product in the collisions,
the weighted average of effective temperatures is nearly the same
as that for pion production. Obviously, in most cases, this
``true" temperature is less than the expected critical temperature
(130--165 MeV) of the QGP formation [37]. In central $p$-Pb
collisions, the ``true" temperature is the closest to the lower
limit of the expected critical temperature.

Comparing with the index $q/(q-1)$, the index $1/(q-1)$ gives a
smaller $q$. Combining with our previous work [36], in the case of
using the same index, we learn that $q$ in Pb-Pb collisions is
less than that in $p$-Pb collisions. If $q=1$ corresponds to the
equilibrium state, the interacting systems in both $p$-Pb and
Pb-Pb collisions are close to the equilibrium state, and the
interacting system in Pb-Pb collisions is closer to the
equilibrium state. At the same time, the interacting systems in
central collisions are closer to the equilibrium state, and the
interacting system consisted of heavier particles is closer to the
equilibrium state, too. These results can be explained by the
larger interacting region (volume) in central collisions and
shorter mean free path of heavier particles.

From Fig. 3 and Table 2, we can see the slopes and intercept
points in different linear correlations. The absolute slopes
$|a|$'s in $T-C$, $q-C$, $T-m_0$, and $q-m_0$ correlations are
less than those in $T-q$ correlations, which renders slow changes
in the former cases. The intercept points $b$'s in $q-C$ and
$q-m_0$ correlations are greater than 1 due to the limitation of
physics condition. In $T-C$ and $T-m_0$ correlations, $b$'s
indicate the maximum (122--311 MeV) and minimum (66--91 MeV)
temperatures respectively. In $q-C$ and $q-m_0$ correlations,
$b$'s indicate the minimum (1.06--1.13) and maximum (1.14--1.15)
non-equilibrium degrees respectively. In $T-q$ correlations, $b$'s
have no physics meaning, because these cases correspond to $q=0$
which is beyond the limitation of $q\geq1$. From $T-q$
correlations one can see that $q$ has a value of 1.12--1.21 if the
interacting system becomes very cool ($T\longrightarrow 0$).

The values of $T$ and $q$ obtained from the above transverse
momentum distributions can be used in analysis of pseudorapidity
distributions in the same or similar conditions. Fig. 4(a)
presents the pseudorapidity distribution of charged particles
produced in NSD $p$-Pb collisions at $\sqrt{s_{NN}}=5.02$ TeV in
the laboratory reference frame, where $N_{ch}$ on the axis denotes
the number of charged particles. The circles represents the
experimental data of the ALICE Collaboration [31]. The dotted and
dashed curves are the contributions of (left target) $p$-cylinder
and (right projectile) Pb-cylinder respectively, and the solid
curve is the sum of the two cylinders in which each source is
described by the Tasllis statistics. The contributions of leading
nucleon sources are neglected. In the calculation, we have
distinguished the pseudorapidity and rapidity, and taken the
weighted average values of $T=93\pm4$ MeV, $q=1.142\pm0.002$, and
$m_0=174\pm2$ MeV/$c^2$ from Fig. 1. Other parameter values
obtained by fitting the experimental pseudorapidity distribution
are $y_{T\min}=-2.65\pm0.08$, $y_{T\max}=0.01\pm0.01$,
$y_{P\min}=0.01\pm0.01$, $y_{P\max}=3.79\pm0.10$, and the
contribution ratio of target cylinder $K_T=0.393\pm0.002$, with
$\chi^2$/dof $=0.046$. One can see that the modelling result is in
agreement with the experimental data in the available $\eta$
range.

Comparing with the pseudorapidity distribution in Fig. 4(a), we
give correspondingly the rapidity distribution of charged
particles produced in NSD $p$-Pb collisions at
$\sqrt{s_{NN}}=5.02$ TeV in Fig. 4(b), where the meanings of
different curves are the same as those in Fig. 4(a). One can see
the difference and similarity between the pseudorapidity and
rapidity distributions. By using the same set of parameter values,
the distributions of transverse rapidity and rapidities in $ox$
($oy$) axis direction for charged particles produced in NSD $p$-Pb
collisions at $\sqrt{s_{NN}}=5.02$ TeV are given in Figs. 4(c) and
4(d) respectively, where the meanings of different curves are the
same as those in Fig. 4(a). One can see that the difference
between the contributions of $p$-cylinder and Pb-cylinder in small
$y_T$ (or $|y_{1,2}|$) region is large, and the difference between
the two contributions in middle-large $y_T$ (or $|y_{1,2}|$)
region is small.

The structure pictures of interacting system (the dispersion plots
of final-state particles) in NSD $p$-Pb collisions at
$\sqrt{s_{NN}}=5.02$ TeV at the stage of kinetic freeze-out in the
rapidity spaces (a) $y_2-y_1$, (b) $y_{1,2}-y_T$, (c) $y_{1,2}-y$,
and (d) $y_T-y$ are given in Fig. 5. The circles and squares
correspond to the contributions of $p$-cylinder and Pb-cylinder
respectively, where the contributions of leading nucleons are not
included. The simulated total number of particles is 1000.
Correspondingly, the simulated numbers of particles produced in
$p$-cylinder and Pb-cylinder are 393 and 607 respectively, due to
different contribution ratios of the two cylinders. One can see
that the densities in small $|y_{1,2}|$ and $y_T$ regions are
larger than those in large $|y_{1,2}|$ and $y_T$ regions. There
are some zero density regions in the rapidity spaces due to the
limitations of kinetics.

The structure pictures of interacting system in NSD $p$-Pb
collisions at $\sqrt{s_{NN}}=5.02$ TeV at the stage of kinetic
freeze-out in the momentum spaces (a) $p_y-p_x$, (b)
$p_{x,y}-p_T$, (c) $p_{x,y}-p_z$, and (d) $p_T-p_z$ are presented
in Fig. 6. The meanings of the symbols are the same as those in
Fig. 5. One can see that the densities in small $|p_{x,y,z}|$ and
$p_T$ regions are larger than those in large $|p_{x,y,z}|$ and
$p_T$ regions.

The structure pictures of interacting system in NSD $p$-Pb
collisions at $\sqrt{s_{NN}}=5.02$ TeV at the stage of kinetic
freeze-out in the velocity spaces (a) $\beta_y-\beta_x$, (b)
$\beta_{x,y}-\beta_T$, (c) $\beta_{x,y}-\beta_z$, and (d)
$\beta_T-\beta_z$ are presented in Fig. 7. Meanwhile, the figure
is also the structure pictures of interacting system in the same
collisions at the stage of kinetic freeze-out in the coordinate
space over $t_0$: (a) $y/t_0-x/t_0$, (b) $x/t_0(y/t_0)-r_T/t_0$,
(c) $x/t_0(y/t_0)-z/t_0$, and (d) $r_T/t_0-z/t_0$. The units of
quantities in the axes are $c$, where $c=1$ in the natural units.
The meanings of the symbols are the same as those in Fig. 5. One
can see that the densities in small $|\beta_{x,y}|$ and $\beta_T$
regions and large $|\beta_z|$ region are larger than those in
large $|\beta_{x,y}|$ and $\beta_T$ regions and small $|\beta_z|$
region. Because all the maximum velocities in different directions
are close to $c$, the structure picture of the interacting system
in the coordinate space is in fact a sphere which has high
densities in near surface regions towards the two beam directions.
\\

{\section{Conclusions}}

From the above discussions, we obtain following conclusions.

(a) The transverse momentum distributions of final-state particles
produced in $p$-Pb collisions at LHC energy can be described by
the Tsallis distribution which reflects the multiple temperature
emission in the multisource thermal model. The calculated results
are in agreement with the experimental data of $\pi^{\pm}$,
$K^{\pm}$, $p+\bar p$, and $\Lambda+\bar \Lambda$ measured by the
ALICE and CMS Collaborations in $p$-Pb collisions with different
centrality intervals at $\sqrt{s_{NN}}=5.02$ TeV.

(b) The Tsallis transverse momentum distribution uses two free
parameters $T$ and $q$ to describe the average temperature and the
non-equilibrium degree of the interacting system respectively. The
physics condition gives $q \geq 1$. A large $q$ corresponds to a
state departing far from equilibrium and $q=1$ corresponds to an
equilibrium state. The present work shows that the values of $q$
are not too large in most cases. This means that the whole
interacting system in $p$-Pb collisions at $\sqrt{s_{NN}}=5.02$
TeV is close to an equilibrium sate. The obtained temperature is
less than the lower limit of the expected critical temperature of
the QGP formation.

(c) The two parameters ($T$ and $q$) depend on the impacting
centrality and particle mass, and there is a correlation between
the two parameters. The present work shows that $T$ decreases and
$q$ increases with increase of the centrality percentage, and $T$
increases and $q$ decreases with increase of the particle mass. A
negative correlation exists between $T$ and $q$ not only for
different centralities but also for different particles in the
mentioned collisions.

(d) The pseudorapidity distribution of charged particles produced
in NSD $p$-Pb collisions at $\sqrt{s_{NN}}=5.02$ TeV can be
described by the multisource thermal model in which each source is
described by the Tsallis statistics. The contributions of
$p$-cylinder and Pb-cylinder are given. The parameter values
obtained by fitting the transverse momentum and pseudorapidity
distributions are used to extract the distributions of rapidities
$y$, transverse rapidities $y_T$, rapidities $y_1$ in $ox$ axis
direction, and rapidities $y_2$ in $oy$ axis direction. The
contributions of the two cylinders are obviously different in the
small $y_T$ or $|y_{1,2}|$ region, and the two contributions are
similar in the middle-large $y_T$ or $|y_{1,2}|$ region.

(e) The structure pictures of interacting system in NSD $p$-Pb
collisions at $\sqrt{s_{NN}}=5.02$ TeV at the stage of kinetic
freeze-out in rapidity spaces are extracted. These structure
pictures are also the dispersion plots of final-state particles in
rapidity spaces. The contributions of $p$-cylinder and Pb-cylinder
are given. The densities in small $|y_{1,2}|$ and $y_T$ regions
are larger than those in large $|y_{1,2}|$ and $y_T$ regions.
There are some zero density regions in the rapidity spaces due to
the limitations of kinetics.

(f) The structure pictures of interacting system in NSD $p$-Pb
collisions at $\sqrt{s_{NN}}=5.02$ TeV at the stage of kinetic
freeze-out in momentum spaces are extracted. These structure
pictures are also the dispersion plots of final-state particles in
momentum spaces. The contributions of $p$-cylinder and Pb-cylinder
are given. The densities in small $|p_{x,y,z}|$ and $p_T$ regions
are larger than those in large $|p_{x,y,z}|$ and $p_T$ regions.
There are some zero density regions in the momentum spaces due to
the limitations of kinetics.

(g) The structure pictures of interacting system in NSD $p$-Pb
collisions at $\sqrt{s_{NN}}=5.02$ TeV at the stage of kinetic
freeze-out in velocity (coordinate) spaces are extracted. These
structure pictures are also the dispersion plots of final-state
particles in velocity (coordinate) spaces. The contributions of
$p$-cylinder and Pb-cylinder are given. The densities in small
$|\beta_{x,y}|$ and $\beta_T$ regions and large $|\beta_z|$ region
are larger than those in large $|\beta_{x,y}|$ and $\beta_T$
regions and small $|\beta_z|$ region. The structure picture of the
interacting system in the coordinate space is in fact a sphere
which has high densities in near surface regions towards the two
beam directions.
\\

{\bf Acknowledgment}

This work was supported by the National Natural Science Foundation
of China under Grant No. 10975095, the Open Research Subject of
the Chinese Academy of Sciences Large-Scale Scientific Facility
under Grant No. 2060205, the Shanxi Provincial Natural Science
Foundation under Grant No. 2013021006, the Shanxi Scholarship
Council of China under Grant No. 2012-012, and the US DOE under
contract DE-FG02-87ER40331.A008.

\newpage

\renewcommand{\baselinestretch}{1.0}

{\small {Table 1. Values of free parameters, normalization
constant, and $\chi^2$/dof corresponding to the curves in Figs. 1
and 2. The relative errors for $T$, $q$, and $C_0$ are around 5\%,
0.2\%, and 5\%, respectively. \\[-1mm]
{%
\begin{center}
\begin{tabular}{ccccccc}
\hline\hline  Figure & Particle & Centrality & $T$ (GeV) & $q$ & $C_0$ (fm$^3$) & $\chi^2$/dof \\
\hline
1(a) & $\pi^+$                & 0-100\%  & 0.08 & 1.151 & 204.8 & 0.141 \\
     & $K^+$                  & 0-100\%  & 0.21 & 1.059 &  42.4 & 0.044 \\
     & $p$                    & 0-100\%  & 0.25 & 1.067 &  37.1 & 0.059 \\
1(b) & $\pi^-$                & 0-100\%  & 0.08 & 1.151 & 204.8 & 0.097 \\
     & $K^-$                  & 0-100\%  & 0.21 & 1.059 &  42.4 & 0.039 \\
     & $\bar p$               & 0-100\%  & 0.25 & 1.067 &  37.1 & 0.052 \\
2(a) & $\pi^{\pm}$            & 0-5\%    & 0.12 & 1.126 &  16.1 & 0.333 \\
     &                        & 5-10\%   & 0.12 & 1.127 &  14.9 & 0.298 \\
     &                        & 10-20\%  & 0.11 & 1.134 &  18.7 & 0.227 \\
     &                        & 20-40\%  & 0.11 & 1.131 &  18.2 & 0.348 \\
     &                        & 40-60\%  & 0.09 & 1.140 &  34.8 & 0.273 \\
     &                        & 60-80\%  & 0.08 & 1.144 &  48.0 & 0.314 \\
     &                        & 80-100\% & 0.07 & 1.142 &  73.6 & 0.339 \\
2(b) & $K^{\pm}$              & 0-5\%    & 0.18 & 1.110 &  18.3 & 0.068 \\
     &                        & 5-10\%   & 0.17 & 1.114 &  21.5 & 0.086 \\
     &                        & 10-20\%  & 0.16 & 1.122 &  23.7 & 0.072 \\
     &                        & 20-40\%  & 0.16 & 1.115 &  25.6 & 0.099 \\
     &                        & 40-60\%  & 0.13 & 1.125 &  50.4 & 0.132 \\
     &                        & 60-80\%  & 0.12 & 1.125 &  64.0 & 0.174 \\
     &                        & 80-100\% & 0.10 & 1.123 & 127.3 & 0.173 \\
2(c) & $p+\bar p$             & 0-5\%    & 0.30 & 1.066 &  11.2 & 0.031 \\
     &                        & 5-10\%   & 0.29 & 1.065 &  13.0 & 0.037 \\
     &                        & 10-20\%  & 0.25 & 1.078 &  19.3 & 0.015 \\
     &                        & 20-40\%  & 0.25 & 1.072 &  21.0 & 0.032 \\
     &                        & 40-60\%  & 0.20 & 1.082 &  49.7 & 0.111 \\
     &                        & 60-80\%  & 0.15 & 1.092 & 156.8 & 0.059 \\
     &                        & 80-100\% & 0.11 & 1.096 & 678.4 & 0.139 \\
2(d) & $\Lambda+\bar \Lambda$ & 0-5\%    & 0.31 & 1.062 &  17.6 & 0.134 \\
     &                        & 5-10\%   & 0.30 & 1.065 &  19.4 & 0.142 \\
     &                        & 10-20\%  & 0.28 & 1.069 &  22.8 & 0.072 \\
     &                        & 20-40\%  & 0.24 & 1.078 &  40.3 & 0.075 \\
     &                        & 40-60\%  & 0.20 & 1.086 &  79.8 & 0.069 \\
     &                        & 60-80\%  & 0.16 & 1.094 & 211.2 & 0.123 \\
     &                        & 80-100\% & 0.13 & 1.096 & 592.6 & 0.354 \\
\hline\hline
\end{tabular}%
\end{center}
}} }

\newpage

{\small {Table 2. Values of coefficients and $\chi^2$/dof
corresponding to the lines in Fig. 3. The units of $T$
and $m_0$ are GeV and GeV/$c^2$ respectively. \\[-1mm]
{%
\begin{center}
\begin{tabular}{cccccc}
\hline\hline  Figure & Correlation & Type & $a$ & $b$ & $\chi^2$/dof \\
\hline
3(a) & $T-C$    & $\pi^{\pm}$            & $-0.059\pm0.004$ & $0.122\pm0.002$ & 0.020 \\
     &          & $K^{\pm}$              & $-0.087\pm0.006$ & $0.179\pm0.003$ & 0.024 \\
     &          & $p+\bar p$             & $-0.212\pm0.013$ & $0.302\pm0.007$ & 0.039 \\
     &          & $\Lambda+\bar \Lambda$ & $-0.210\pm0.008$ & $0.311\pm0.004$ & 0.029 \\
3(b) & $q-C$    & $\pi^{\pm}$            &  $0.020\pm0.004$ & $1.127\pm0.002$ & 0.004 \\
     &          & $K^{\pm}$              &  $0.014\pm0.005$ & $1.114\pm0.003$ & 0.008 \\
     &          & $p+\bar p$             &  $0.035\pm0.005$ & $1.066\pm0.002$ & 0.007 \\
     &          & $\Lambda+\bar \Lambda$ &  $0.041\pm0.003$ & $1.063\pm0.002$ & 0.003 \\
3(c) & $T-m_0$  & 0-5\%                  &  $0.208\pm0.016$ & $0.088\pm0.012$ & 0.084 \\
     &          & 5-10\%                 &  $0.199\pm0.017$ & $0.086\pm0.013$ & 0.139 \\
     &          & 10-20\%                &  $0.178\pm0.007$ & $0.081\pm0.005$ & 0.047 \\
     &          & 20-40\%                &  $0.147\pm0.018$ & $0.091\pm0.014$ & 0.121 \\
     &          & 40-60\%                &  $0.122\pm0.011$ & $0.073\pm0.009$ & 0.068 \\
     &          & 60-80\%                &  $0.081\pm0.006$ & $0.073\pm0.005$ & 0.070 \\
     &          & 80-100\%               &  $0.054\pm0.008$ & $0.066\pm0.254$ & 0.006 \\
3(d) & $q-m_0$  & 0-5\%                  & $-0.071\pm0.007$ & $1.139\pm0.005$ & 0.019 \\
     &          & 5-10\%                 & $-0.071\pm0.009$ & $1.141\pm0.007$ & 0.038 \\
     &          & 10-20\%                & $-0.071\pm0.007$ & $1.149\pm0.006$ & 0.022 \\
     &          & 20-40\%                & $-0.062\pm0.009$ & $1.141\pm0.007$ & 0.038 \\
     &          & 40-60\%                & $-0.063\pm0.009$ & $1.150\pm0.007$ & 0.033 \\
     &          & 60-80\%                & $-0.056\pm0.006$ & $1.151\pm0.254$ & 0.005 \\
     &          & 80-100\%               & $-0.050\pm0.004$ & $1.148\pm0.003$ & 0.009 \\
3(e) & $T-q$    & $\pi^{\pm}$            & $-2.625\pm0.378$ & $3.078\pm0.429$ & 0.130 \\
     &          & $K^{\pm}$              & $-4.013\pm1.236$ & $4.637\pm1.384$ & 0.424 \\
     &          & $p+\bar p$             & $-5.780\pm0.478$ & $6.457\pm0.516$ & 0.101 \\
     &          & $\Lambda+\bar \Lambda$ & $-5.056\pm0.235$ & $5.684\pm0.254$ & 0.052 \\
3(f) & $T-q$    & 0-5\%                  & $-2.907\pm0.095$ & $3.399\pm0.104$ & 0.042 \\
     &          & 5-10\%                 & $-2.734\pm0.119$ & $3.207\pm0.130$ & 0.056 \\
     &          & 10-20\%                & $-2.439\pm0.154$ & $2.885\pm0.170$ & 0.125 \\
     &          & 20-40\%                & $-2.330\pm0.104$ & $2.751\pm0.114$ & 0.046 \\
     &          & 40-60\%                & $-1.887\pm0.107$ & $2.246\pm0.119$ & 0.065 \\
     &          & 60-80\%                & $-1.392\pm0.165$ & $1.678\pm0.184$ & 0.128 \\
     &          & 80-100\%               & $-1.033\pm0.222$ & $1.253\pm0.247$ & 0.218 \\
\hline\hline
\end{tabular}%
\end{center}
}} }

\newpage
\begin{figure}
\hskip-1.0cm \begin{center}
\includegraphics[width=16.0cm]{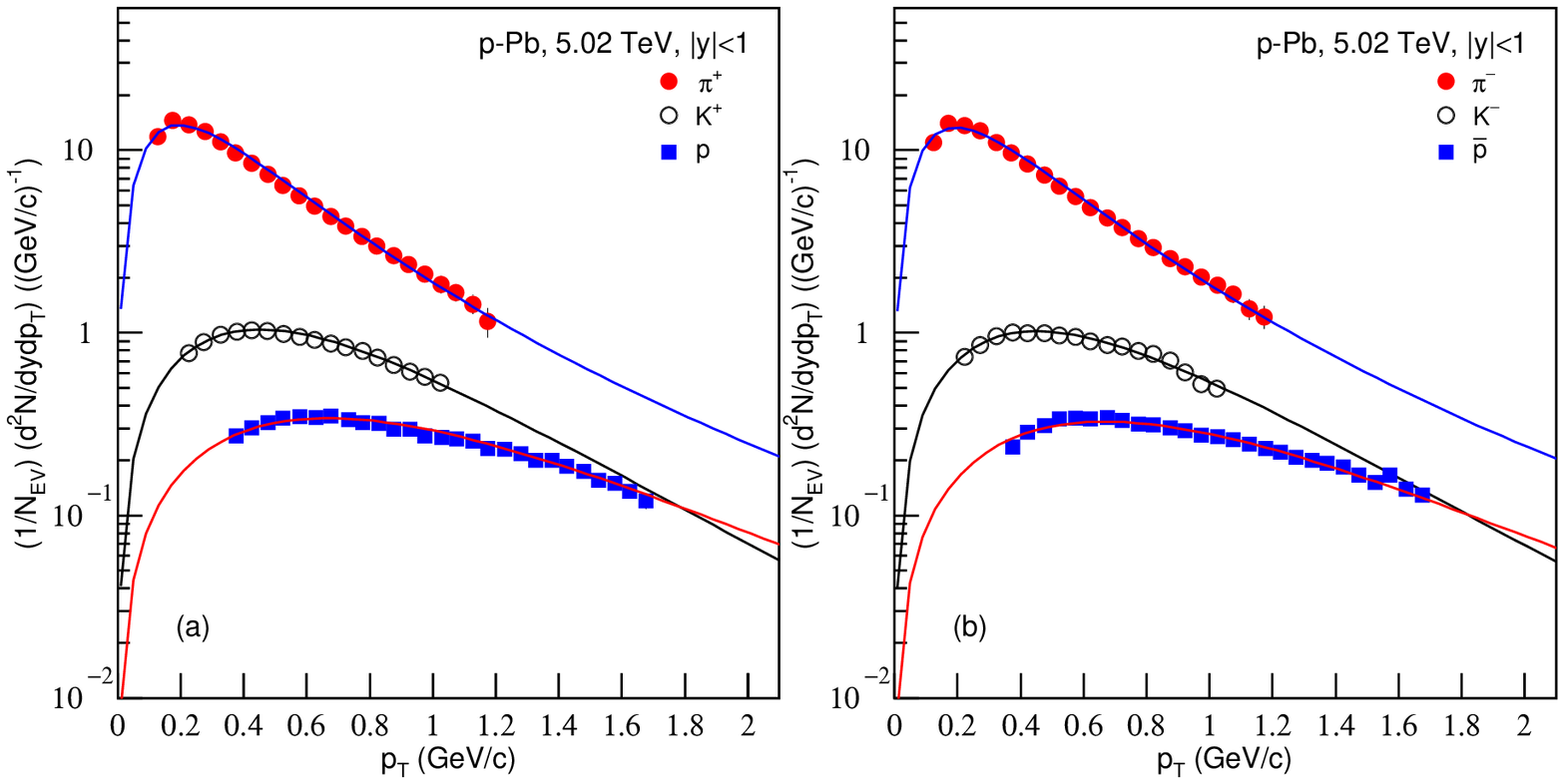}
\end{center}
\vskip1.0cm Fig. 1. Transverse momentum distributions of (a)
$\pi^+$, $K^+$, and $p$, as well as (b) $\pi^-$, $K^-$, and $\bar
p$ produced in $p$-Pb collisions at $\sqrt{s_{NN}}=5.02$ TeV. The
symbols represent the experimental data of the CMS Collaboration
[32] and the curves are our results calculated by using the
Tsallis transverse momentum distribution.
\end{figure}

\newpage
\begin{figure}
\hskip-1.0cm \begin{center}
\includegraphics[width=16.0cm]{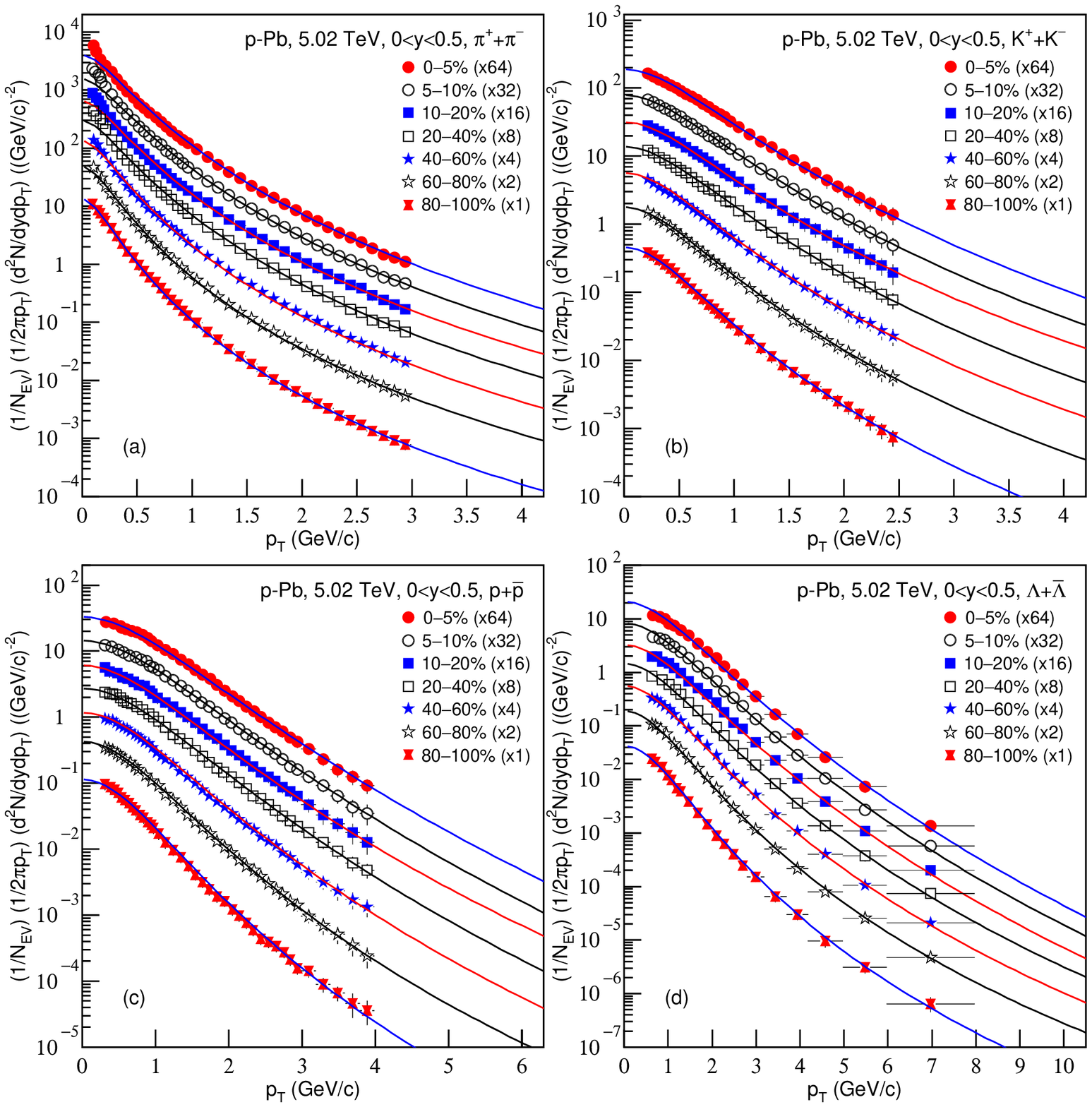}
\end{center}
\vskip1.0cm Fig. 2. Transverse momentum distributions of (a)
$\pi^{\pm}$, (b) $K^{\pm}$, (c) $p+\bar p$, and (d) $\Lambda+\bar
\Lambda$ produced in $p$-Pb collisions with different centrality
intervals at $\sqrt{s_{NN}}=5.02$ TeV. The symbols represent the
experimental data of the ALICE Collaboration [30] and the curves
are our results calculated by using the Tsallis distribution.
\end{figure}

\newpage
\begin{figure}
\hskip-1.0cm \begin{center}
\includegraphics[width=16.0cm]{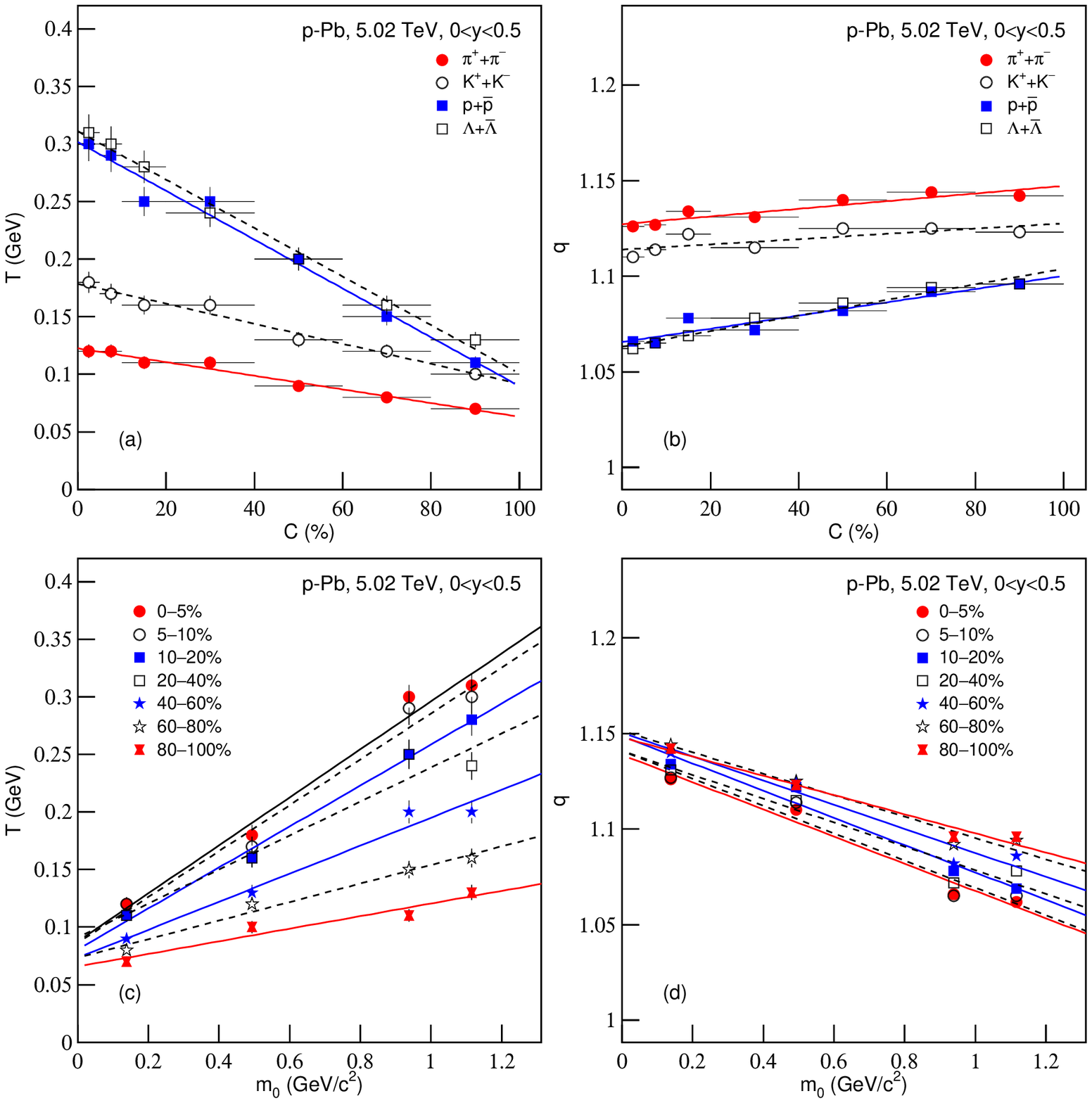}
\end{center}
\vskip1.0cm Fig. 3. The relations (a) $T-C$ for different
particles, (b) $q-C$ for different particles, (c) $T-m_0$ for
different centrality intervals, (d) $q-m_0$ for different
centrality intervals, (e) $T-q$ for different particles, and (f)
$T-q$ for different centrality intervals, in $p$-Pb collisions at
$\sqrt{s_{NN}}=5.02$ TeV. The symbols represent the parameter
values listed in Table 1, and the lines are our fitting results.
\end{figure}

\newpage
\begin{figure}
\hskip-1.0cm \begin{center}
\includegraphics[width=16.0cm]{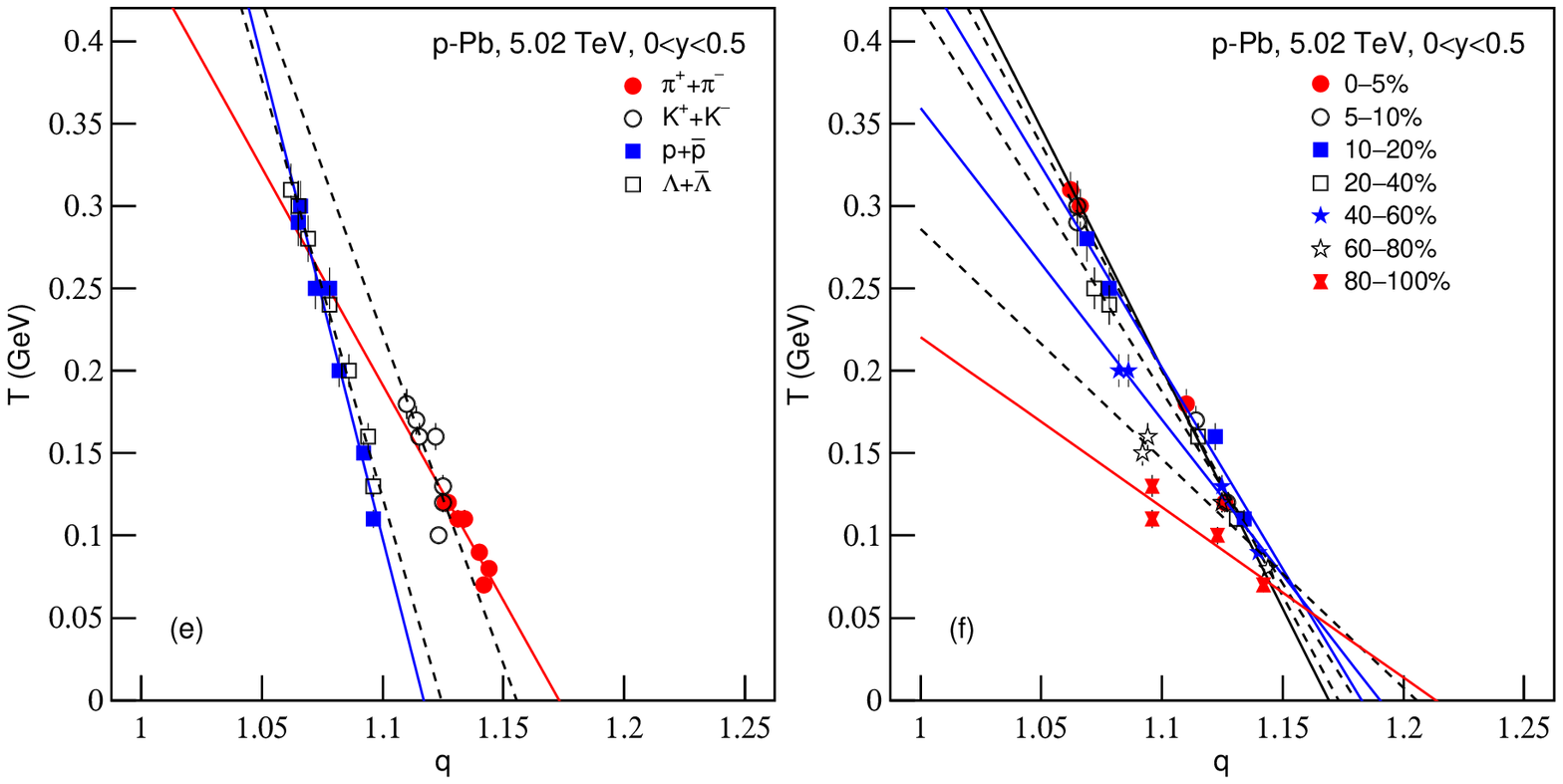}
\end{center}
\vskip1.0cm Fig. 3. Continued.
\end{figure}

\newpage
\begin{figure}
\hskip-1.0cm \begin{center}
\includegraphics[width=16.0cm]{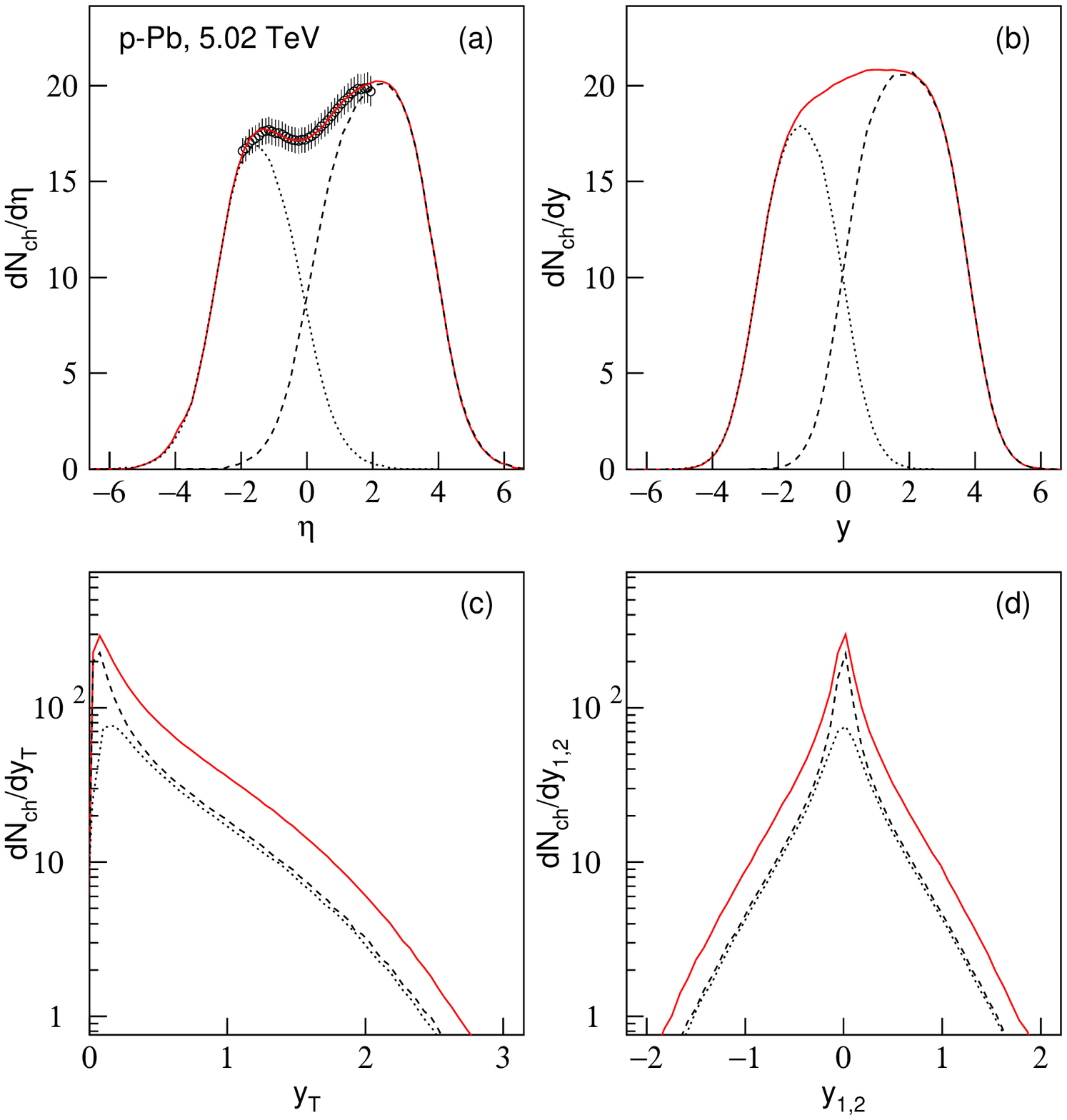}
\end{center}
\vskip1.0cm Fig. 4. (a) Pseudorapidity distribution of charged
particles produced in NSD $p$-Pb collisions at
$\sqrt{s_{NN}}=5.02$ TeV. The circles represents the experimental
data of the ALICE Collaboration [31] and the curves are our
modelling results. The dotted, dashed, and solid curves are the
contributions of $p$-cylinder, Pb-cylinder, and both the
cylinders, respectively. (b)-(d) Correspondingly distributions of
(b) rapidities, (c) transverse rapidities, and (d) rapidities in
$ox$ $(oy)$ axis direction, in the mentioned collisions.
\end{figure}

\newpage
\begin{figure}
\hskip-1.0cm \begin{center}
\includegraphics[width=16.0cm]{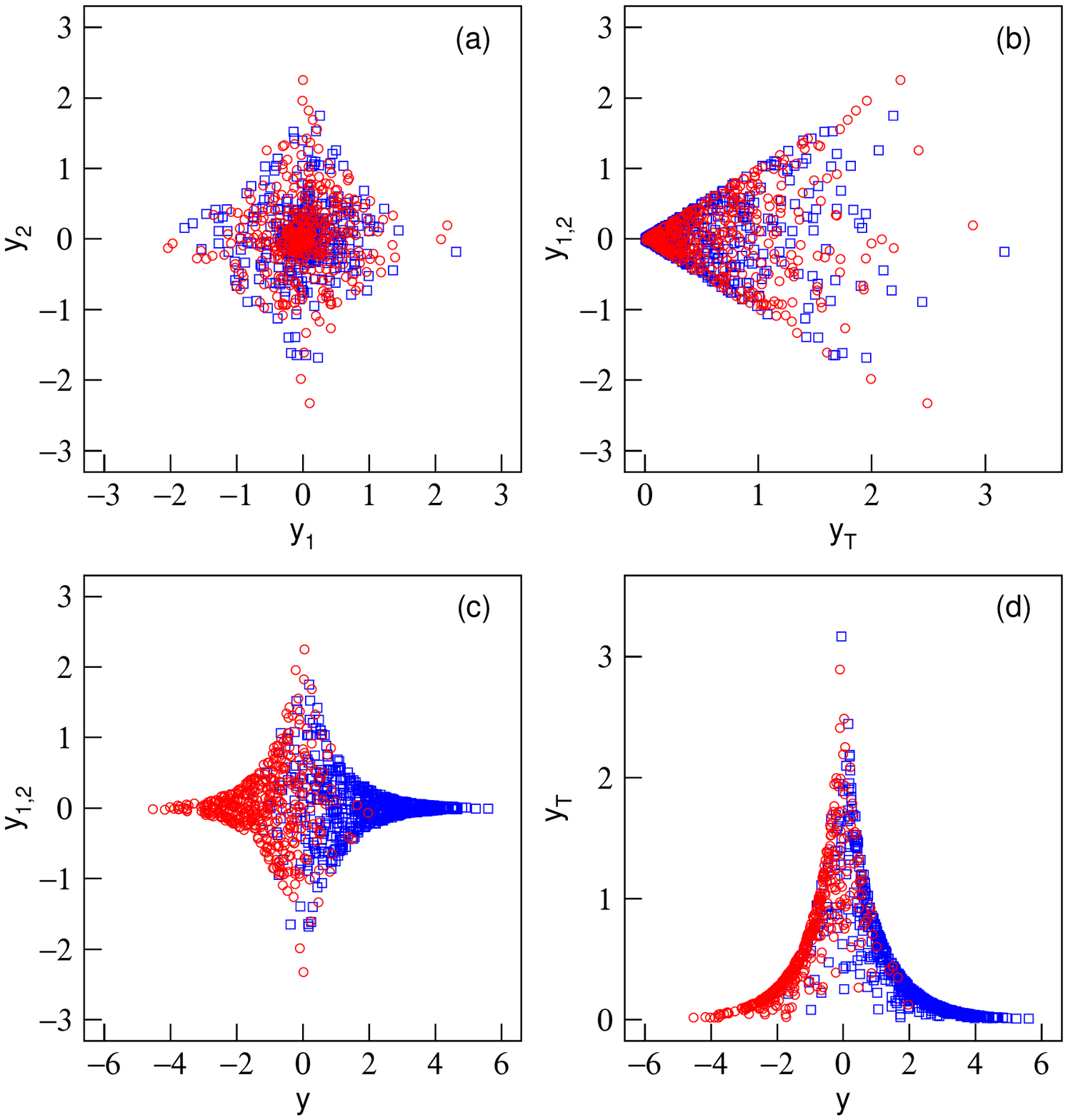}
\end{center}
\vskip1.0cm Fig. 5. Structure pictures of interacting system at
the stage of kinetic freeze-out in the rapidity spaces (a)
$y_2-y_1$, (b) $y_{1,2}-y_T$, (c) $y_{1,2}-y$, and (d) $y_T-y$, in
NSD $p$-Pb collisions at $\sqrt{s_{NN}}=5.02$ TeV. The circles and
squares represent the contributions of $p$-cylinder and
Pb-cylinder respectively. The contributions of leading nucleons
are not included.
\end{figure}

\newpage
\begin{figure}
\hskip-1.0cm \begin{center}
\includegraphics[width=16.0cm]{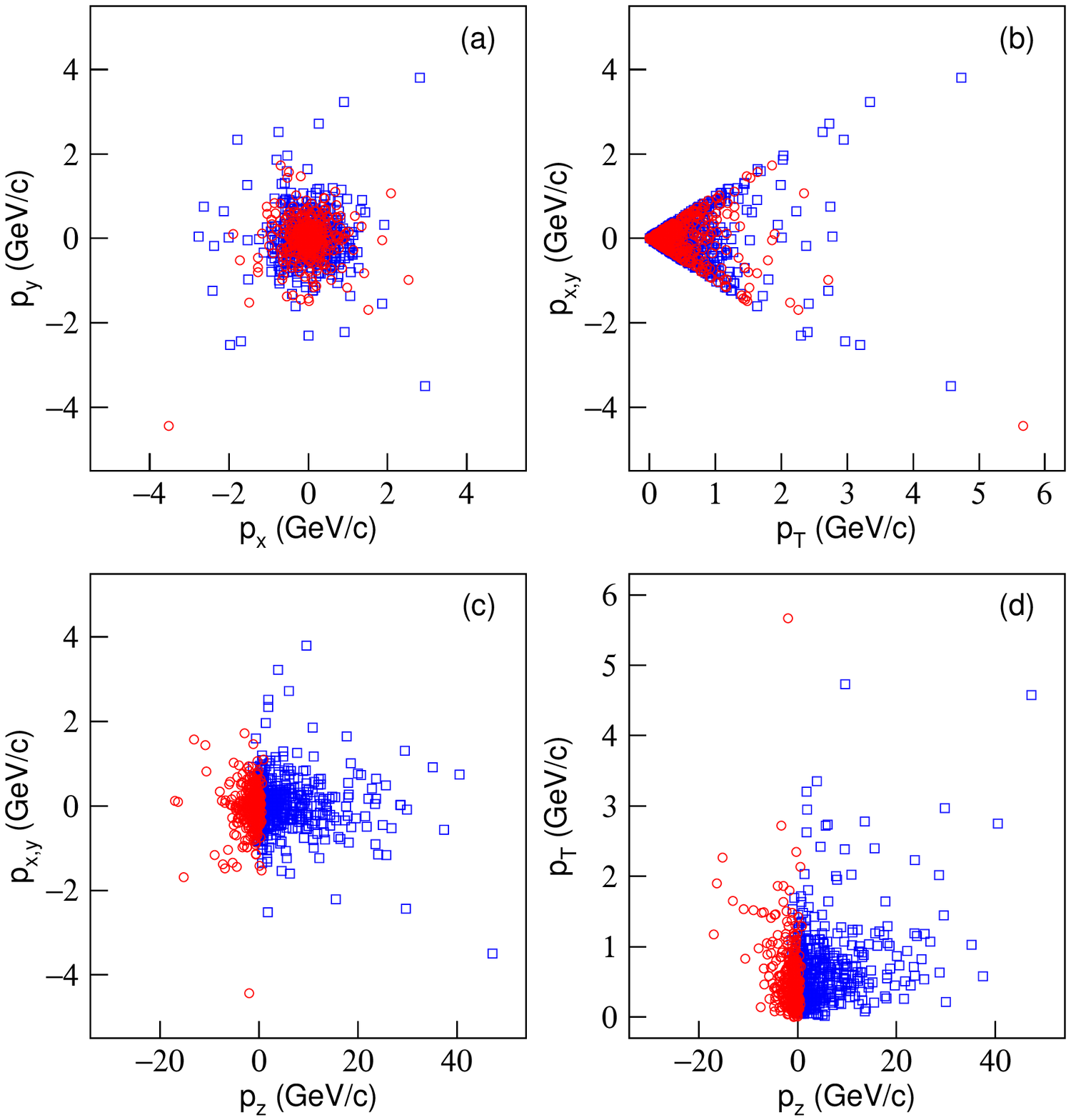}
\end{center}
\vskip1.0cm Fig. 6. As for Fig. 5, but showing the structure
pictures of interacting system at the stage of kinetic freeze-out
in the momentum spaces: (a) $p_y-p_x$, (b) $p_{x,y}-p_T$, (c)
$p_{x,y}-p_z$, and (d) $p_T-p_z$.
\end{figure}

\newpage
\begin{figure}
\hskip-1.0cm \begin{center}
\includegraphics[width=16.0cm]{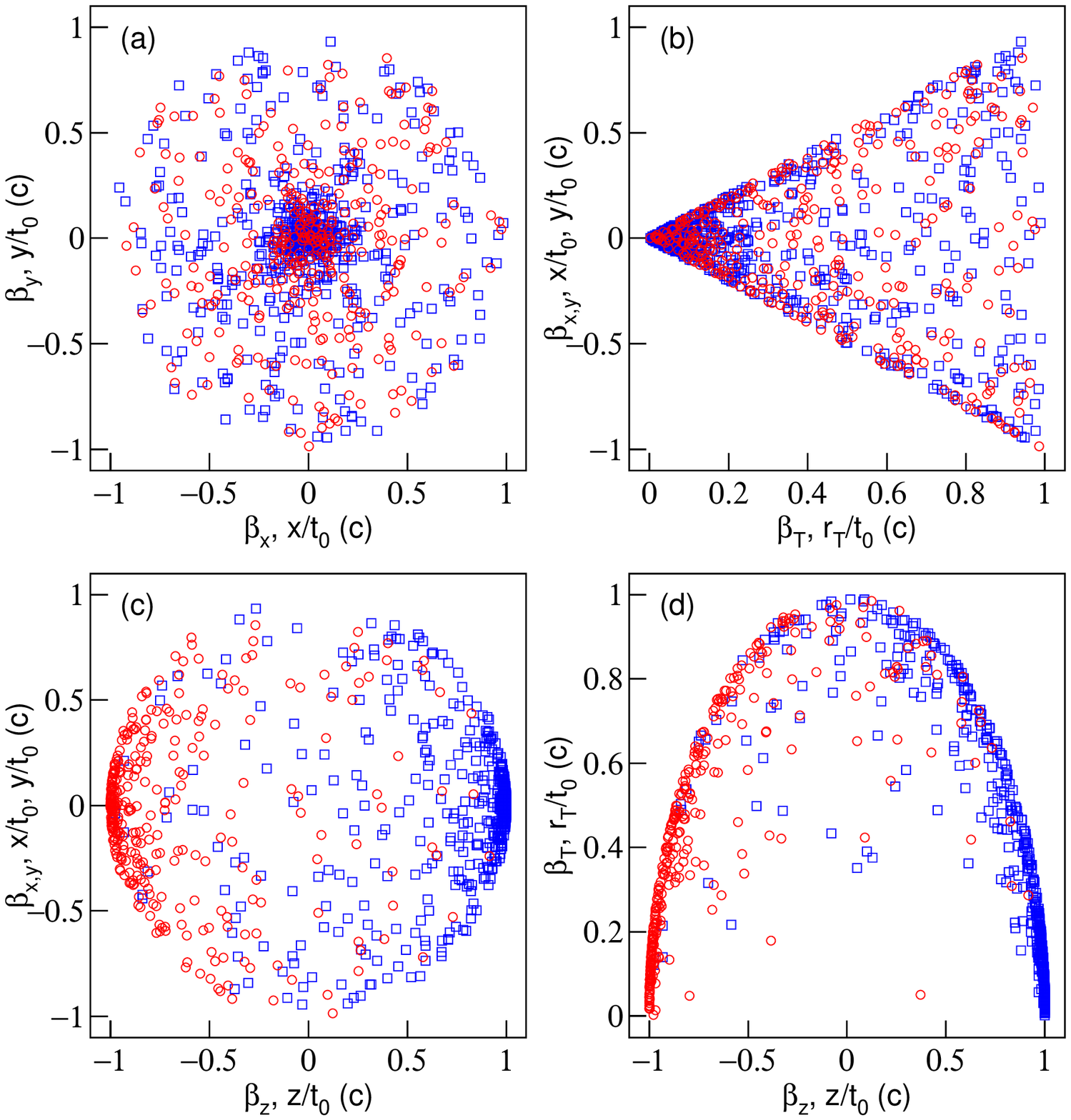}
\end{center}
\vskip1.0cm Fig. 7. As for Fig 5, but showing the structure
pictures of interacting system at the stage of kinetic freeze-out
in the velocity spaces: (a) $\beta_y-\beta_x$, (b)
$\beta_{x,y}-\beta_T$, (c) $\beta_{x,y}-\beta_z$, and (d)
$\beta_T-\beta_z$; or in the coordinate space over $t_0$: (a)
$y/t_0-x/t_0$, (b) $x/t_0(y/t_0)-r_T/t_0$, (c)
$x/t_0(y/t_0)-z/t_0$, and (d) $r_T/t_0-z/t_0$.
\end{figure}

\end{document}